\documentclass[prl,aps,10pt,twocolumn,superscriptaddress,notitlepage,nofootinbib,nobibnotes,amssymb,amsmath]{revtex4-1}
\usepackage{slashed}
\usepackage{mathtools}
\usepackage{bm}
\usepackage{graphicx}
\newcommand{\beqn}{\begin{eqnarray}}
\newcommand{\eeqn}{\end{eqnarray}}
\newcommand{\beqs}{\begin{subequations}}
\newcommand{\eeqs}{\end{subequations}\\[-2mm]\noindent}
\newcommand{\eq}[1]{(\ref{#1})}

\newcommand{\CSW}{{\mbox{\tiny{CSW}}}}

\newcommand{\bs}{\boldsymbol}

\newcommand{\Vector}[2]{{\left( \begin{array}{c} #1 \\ #2 \end{array} \right)}}

\newcommand{\Matrix}[4]{{\left( \begin{array}{cc} #1 & #2 \\ #3 & #4 \end{array} \right)}}

\usepackage{color}
\definecolor{purple}{rgb}{0.8,0,0.6}

\begin{document}

\title{Chiral sound waves in strained Weyl semimetals}

\author{M. N. Chernodub}
\affiliation{Institut Denis Poisson UMR 7013, Universit\'e de Tours, Tours, 37200, France}
\affiliation{Laboratory of Physics of Living Matter, Far Eastern Federal University, Sukhanova 8, Vladivostok, 690950, Russia}

\author{Mar\'ia A. H. Vozmediano}
\affiliation{Instituto de Ciencia de Materiales de Madrid, CSIC, Cantoblanco, Madrid, 28049, Spain.}

\date{December 17, 2019}

\begin{abstract}
We show that a strained wire of a Weyl semimetal supports a new type of gapless excitation, the chiral sound wave (CSW). It is a longitudinal charge density wave analog to the chiral magnetic wave predicted in the quark-gluon plasma but driven by an elastic axial pseudo-magnetic field. It involves the axial-axial-axial contribution to the chiral anomaly which couples the chiral charge density to the elastic axial gauge field. The chiral sound is unidirectional: it propagates along the elastic magnetic field and not in the opposite direction. The CSW may propagate for long distances as it does not couple directly to quickly dissipating electromagnetic plasmons, while its damping is controlled by the slow chirality flip rate.  We propose an experimental setup to directly detect the chiral sound, which is excited by mechanical vibrations of the crystal lattice in the GHz frequency range. Our findings contribute to a new trend, the chiral acoustics, in strained Weyl semimetals.
\end{abstract}

\maketitle
The low energy electronic excitations of Dirac and Weyl semimetals in three dimensions are Weyl fermions \cite{Letal14a,Letal14b,NX14,Xu15,Lv15,Xuetal15}. Despite the complexity of the band structure of real materials they have been providing evidences of high energy phenomena often related to quantum anomalies, in particular, experimental evidences for the chiral anomaly \cite{Adler69,BJ69,NN81,VF13,Bur15} and the chiral magnetic effect \cite{FKW08} have been reported in semimetals \cite{XKetal15,Lietal15,HZetal15,ZXetal16,LKetal16}. More recently, the gravitational \cite{Landsteiner:2016led,Getal17}, conformal \cite{Chernodub16,CCV18,ACV19,CV19} and torsional \cite{FKBB19} anomalies have also been incorporated to the play. Directly related to the chiral anomaly there is a prediction in the physics of the quark--gluon plasma, of the existence of a collective excitation called chiral magnetic wave \cite{KY11} which have eluded experimental detection so far due to its strong hybridization with the plasmons \cite{SRG18}. The CMW is a collective, gapless excitation of the electronic fluid which corresponds to a coherent propagation of electric and chiral density waves coupled together by the chiral magnetic effect. In this work we propose  a new type of gapless excitation in strained Weyl semimetals, the chiral sound wave  which can be easier to detect than its magnetic counterpart. 

\paragraph{The chiral sound wave.}
Lattice deformations couple to electronic degrees of freedom of Weyl materials in the form of elastic gauge fields constructed with the deformation tensor. 
The construction of elastic gauge fields in three dimensional (3D) Weyl semimetals (WSM) in~\cite{Cortijo:2016yph} has been followed by a number of works analyzing their physical consequences \cite{PCF16,Cortijo:2016wnf,GVetal16,CZ16,ACV17,LPF17,GMetal17a,GMetal17b,GMetal17c,GMSS17,AV18}. An experimental realization of elastic gauge fields in a WSM has appeared recently \cite{KS19}. The most interesting feature of these elastic gauge fields is that they couple to the two chiralities with opposite signs, {\it i. e.}, they are axial pseudo-gauge fields.

In a general  case, for a Weyl semimetal with two Weyl nodes separated in momentum space by a vector ${\bs b}$, the  axial gauge field induced by an elastic deformation of the lattice described by the displacement vector ${\bs u}$ is given, in a simplified form, by the following equation~\cite{Cortijo:2016yph}
\beqn
A_i^5 = \beta u_{ij} b^j, 
\label{eq:AEB:5}
\eeqn
where $u_{ij}$ is the  strain tensor~\cite{LL71b}:
\beqn
u_{ij}(x) = \frac{1}{2}(\partial_i u_j+\partial_ju_i),
\label{eq:uij}
\eeqn
and $\beta$ is a Gruneisen parameter. The axial electric and magnetic fields are defined in the standard way, respectively: $B^{i5}=\frac{1}{2}\epsilon^{ijk}\partial_jA_k^5$, $E_i^{5}=-\partial_t A_i$. To simplify notations, we work in units $\hbar = c = 1$.

\begin{figure}[!thb]
\begin{center}
\includegraphics[scale=0.2,clip=true]{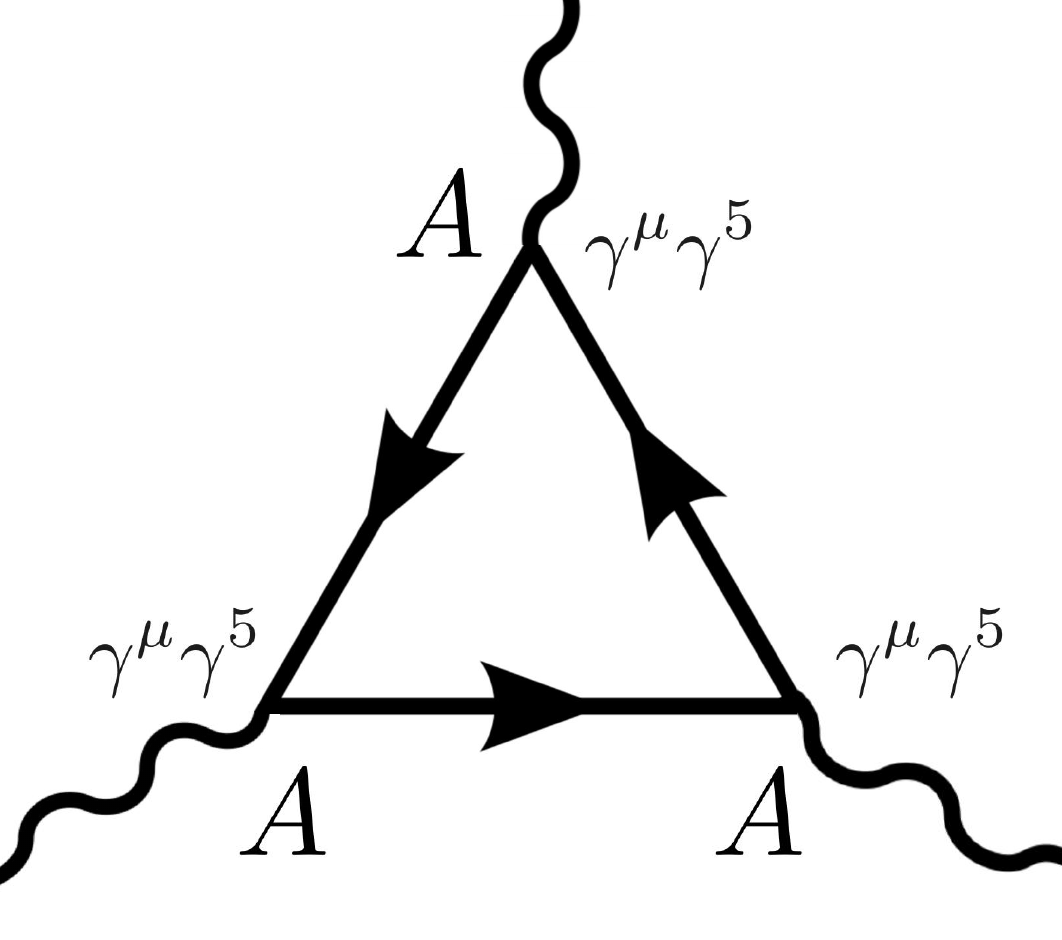} 
\end{center}
\vskip -5mm
\caption{The triangular diagram for the AAA anomaly~\eq{eq:AAA}.}
\label{fig:AAA}
\end{figure}

These fields lead to the nonconservation of the axial charge via axial-axial-axial (AAA) triangle anomaly~\cite{Landsteiner:2016led}:
\beqn
\partial_\mu j^\mu_5 = \frac{1}{24 \pi^2} F_{5,\mu\nu} {\widetilde F}_5^{\mu\nu} \equiv \frac{1}{3}\cdot\frac{1}{ 2 \pi^2} {\bs E}_5 \cdot {\bs B}_5,
\label{eq:AAA}
\eeqn
where $ {\widetilde F}_5^{\mu\nu}  = 1/2 \epsilon^{\mu\nu\alpha\beta} F_{5,\alpha\beta}$ and $j^\mu_5 = j^\mu_L - j^\mu_R$ is the axial current. The corresponding diagram is shown in Fig.~\ref{fig:AAA}.

In the presence of a chiral imbalance, the AAA triangle anomaly~\eq{eq:AAA} leads to the generation of an axial current~${\bs j}_5$ along the direction of the axial magnetic field~${\bs B}_5$:
\beqn
{\bs j}_5 = \frac{\mu_5}{2 \pi^2} {\bs B}_5,
\label{eq:j5}
\eeqn
where $\mu_5 = (\mu_L - \mu_R)/2$ is the chiral (axial) chemical potential that encodes the difference in the Fermi levels at the left-handed ($\mu_L$) and right-handed ($\mu_R$) Weyl cones. The usual (vector) chemical potential is $\mu = (\mu_L + \mu_R)/2$.

Let us fix the axial magnetic field along the $z$ axis, ${\bs B}_5 = (0,0,B^z_5)$ and consider the propagation of the axial current along the magnetic field, ${\bs j}_5 = (0,0,j_5^z)$. For a static stress, the axial electric field vanishes, ${\bs E}_5 = 0$, and the axial charge in Eq.~\eq{eq:AAA} is conserved:
\beqn
\frac{\partial \rho_5}{\partial t} + \frac{\partial j_5^z}{\partial z} = 0,
\label{eq:j5:cons}
\eeqn
where $\rho_5 \equiv j_5^0$ is the chiral charge density. For weak axial-magnetic field background fields, $|B_5| \ll \min(T^2, \mu^2)/v_F^2$, the chiral density is determined, in thermodynamic equilibrium, by the chemical potentials $\mu$ and $\mu_5$, the temperature $T$, and the Fermi velocity~$v_F$:
\beqn
\rho_5 = \frac{\mu_5}{3 v_F^3} \left(T^2 + \frac{3 \mu^2}{\pi^2} \right) + \frac{\mu_5^3}{3 \pi^2 v_F^3}.
\label{eq:rho:5}
\eeqn

For small perturbations in the axial charge density, $\mu_5 \ll \max (\mu,\pi T)$, the last term in Eq.~\eq{eq:rho:5} may be neglected. Combining Eqs.~\eq{eq:j5} and  \eq{eq:j5:cons} with the linearized Eq.~\eq{eq:rho:5}, we get that the axial-density perturbations in the long-wavelength limit obey the  equation:
\beqn
\left(\frac{\partial}{\partial t} + v_\CSW \frac{\partial}{\partial z}  \right) \rho_5 = 0,
\label{eq:CSW:cons}
\eeqn
which supports gapless solutions $\rho_5(t,z) = f(z - v_\CSW t)$ propagating with velocity
\beqn
v_\CSW = \frac{3 B_5 v_F^3}{2 \left( \pi^2 T^2 + 3 \mu^2 \right)} \ \qquad (\mbox{weak}\ B_5),
\label{eq:CSW}
\eeqn
along the axis of the axial magnetic field. 

In the opposite limit of a strong axial magnetic field, $|B_5| \gg \left[\max(T^2, \mu^2)/v_F^2\right]$, the system enters the  quantum limit where only the lowest Landau level is occupied, and the chiral density is simply 
\beqn
\rho_5 = \frac{|B_5|}{2\pi^2} \frac{\mu_5}{v_F}.
\label{eq:rho:5:strong}
\eeqn
In this case the velocity of the mode~\eq{eq:CSW:cons} equals the Fermi velocity: $v_\CSW = (\mathrm{sign}\,B_5) v_F$.

An important feature of this  chiral sound is that its propagation  is strictly unidirectional: the wave propagates only in the positive direction along the axis of the pseudo magnetic field and not in the opposite way. We call this new gapless mode the {\it Chiral Sound Wave} (CSW). The CSW has a linear dispersion, $\omega = v_\CSW {\bs n} {\bs k}$, where ${\bs n} \equiv {\bs B}_5/B_5$ is the unit vector pointing into the direction of the elastic magnetic field~${\bs B}_5$. 

Our derivation of the CSW follows exactly that of the chiral magnetic wave \cite{KY11} replacing the magnetic field ${\bs B}$ by the elastic pseudomagnetic field ${\bs B}_5$. Hence, our wave does not couple to the electromagnetic fields at linear order and does not mix with the quickly dissipating plasmon modes. The CSW propagates longer and dissipates slower than the chiral magnetic wave \cite{SRG18}.

\paragraph{Experimental setup. Interplay of chiral and  ordinary phonons. }

Consider a long straight rod of length $L$ and radius $R$ made of a  Weyl semimetal. We choose the $z$-axis along the symmetry axis of the rod and twist it uniformly about this direction. The degree of the twist is given by the torsion angle $\theta$ which determines the angle of the rotation of the rod $\varphi$ per its unit length: $\theta = \partial \varphi /\partial z$. For $|\theta| R \ll 1$,  the twist induces the strain $u_x = - \theta y z$, $u_y = \theta x z$, and $u_z = 0$. According to Eq.~\eq{eq:uij}, the strain tensor for the twisted rod has only two nonzero components: $u^{xz} = - \theta y/2$ and $u^{yz} = \theta x/2$. 

\begin{figure}[!thb]
\begin{center}
\vskip -4mm
\includegraphics[scale=0.089,clip=true]{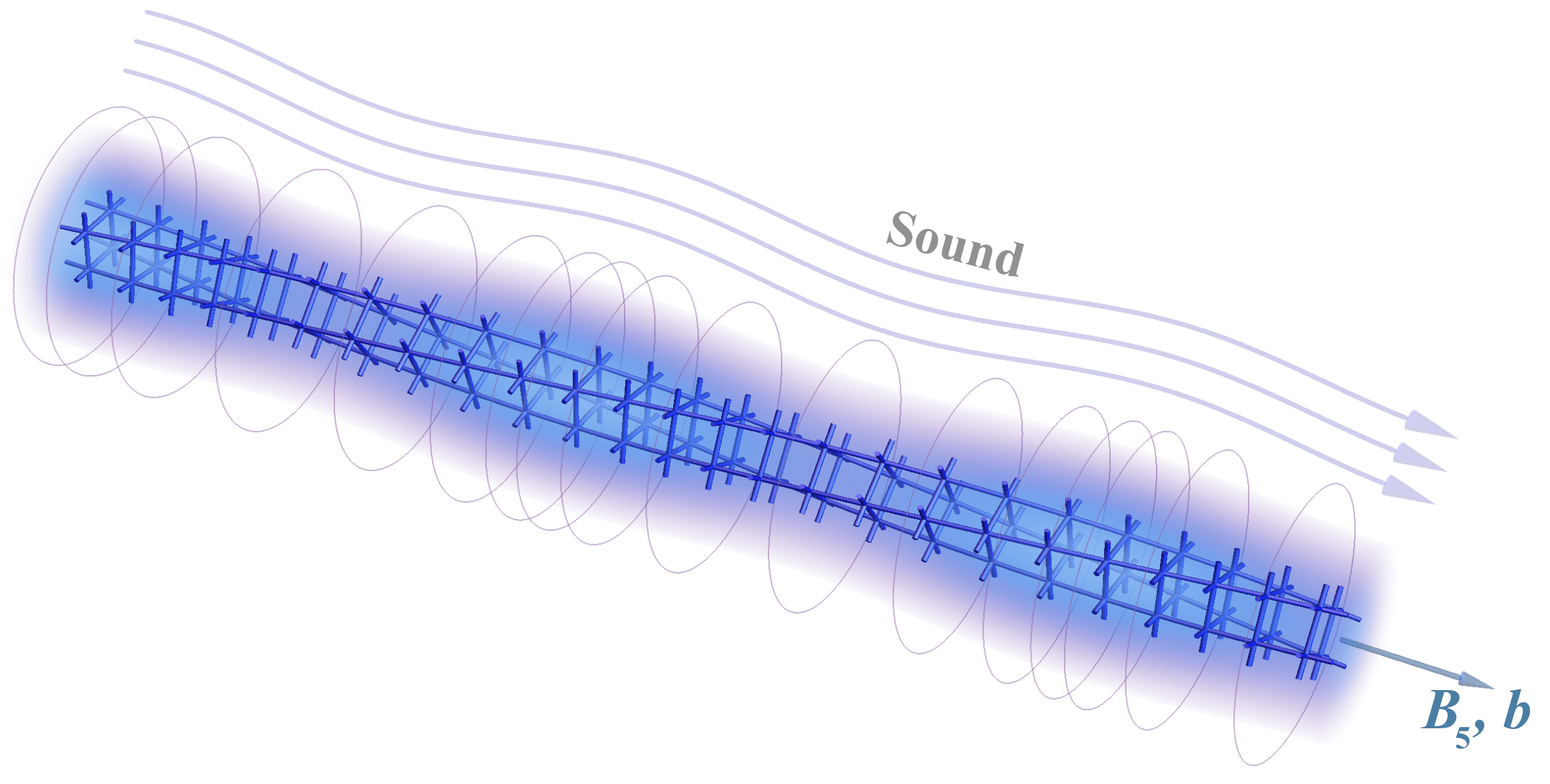} 
\end{center}
\vskip -5mm
\caption{The chiral sound propagates unidirectionally along the axis of the strain-induced pseudo-magnetic ${\bs B}_5$ field.}
\label{fig:setup}
\end{figure}

Consider a rod with the symmetry axis along the direction of the internode vector ${\bs b} \equiv b \,{\mathrm{\bf e}}_z$
as shown in Fig.~\ref{fig:setup}. In this case the elastic gauge field takes the  form ${\bs A}_5 = (- y, x, 0) \theta b/2$, and the corresponding axial pseudomagnetic field is ${\bs B}_5 = \theta {\bs b}$, in the {\it interior} (bulk) of the rod. As it was discussed in~\cite{PCF16}, a flux of the same magnitude and opposite direction will be generated at the surface such that the total flux  through the cross--section of the rod is zero. We discuss only the bulk effects.
 
Longitudinal vibrations of the lattice (standard phonons), will induce an axial electric field parallel to the pseudo--magnetic field thus activating the AAA chiral anomaly in Eq.~\eq{eq:AAA}. In such a way, standard longitudinal phonons will couple with  the chiral phonons. This is the basis of the proposal for the experimental setup to hear the chiral sound due to the elastic deformations of the Weyl crystal. 

The elastic deformation $ u^z (t,{\bs x})$ of the crystal lattice affects the axial current $j^\mu_5$ via Eq.~\eq{eq:AAA}:
\beqn
\frac{\partial \rho_5}{\partial t} + \frac{\partial j_5^z}{\partial z} = - \frac{\theta b^2}{6 \pi^2} \frac{\partial^2 u^z}{\partial t \partial z} - \frac{\rho_5}{\tau_5} 
\label{eq:j5:strain}
\eeqn
where the last term accounts for the relaxation of the chiral charge via inter-valley transitions with the characteristic time $\tau_5$. 

The axial current, in turn, affects the propagation of the ordinary phonons. The electron--phonon interaction is described by the Hamiltonian~\cite{Cortijo:2016yph}
\beqn
H_{int} = \kappa {\bar \psi} {\bs A}_{5} {\bs \gamma} \gamma^5 \psi = \kappa j^z_5 A^z_5  =  \kappa b j^z_5 \frac{\partial u^z}{\partial  z} ,
\label{eq:H:int}
\eeqn
where we have used the explicit form of the elastic field~\eq{eq:AEB:5}. The wave equation for longitudinal phonons propagating along the $z$ axis is given by a variation of a phonon Hamiltonian with respect to the deformation $u^z$:
\beqs
\beqn
\left( \frac{1}{v_s^2} \frac{\partial^2}{\partial t^2} - \frac{\partial^2}{\partial z^2} \right) u^z + \kappa b  \frac{\partial j^z_5}{\partial  z} = 0, 
\label{eq:wave:ph:a}\\
v_s = \sqrt{\frac{3 B + 4 G}{3 \rho}},
\qquad \kappa = \frac{3 \beta}{3B + 4 G}.
\label{eq:wave:ph:b}
\eeqn
\label{eq:wave:ph}
\eeqs
The standard wave equation for the longitudinal sound waves~\eq{eq:wave:ph:a} contains the unusual contribution from the last term originating from the interaction~\eq{eq:H:int} of the elastic deformation with the axial current. Here $v_s \equiv v_L$ is the longitudinal velocity in the absence of the elastic twist, $B$ and $G$ are, respectively, the bulk and shear moduli of the Weyl crystal, and $\rho$ is its density~\cite{LL71b}.

In Eqs.~\eq{eq:j5:strain} and \eq{eq:wave:ph:a}, the $z$ component of the axial current $j^z_5$ is related to the axial charge density $\rho_5$ via the transport law Eq.~\eq{eq:j5} generated by the AAA triangle: 
\beqn
j_5^z = v_\CSW \rho_5\,, 
\qquad
v_\CSW = \frac{3 \theta b v_F^3}{2(\pi^2 T^2 + 3 \mu^2)},
\label{eq:CSW:el}
\eeqn
where $v_\CSW$ is the velocity of the chiral sound wave~\eq{eq:CSW}. Equations~\eq{eq:j5:strain}, \eq{eq:wave:ph}, and \eq{eq:CSW:el} describe the mixed propagation of chiral and ordinary phonons. 

\paragraph{Propagating modes. Sound attenuation. }
Consider the plane wave solutions for both ordinary: $u^z_k(t,z) = u^z_k e^{- i \omega t + i k z }$ and chiral phonons: $\rho_{5,k}(t,z) = \rho_{5,k} e^{- i \omega t + i k z}$, respectively. According to Eqs.~\eq{eq:j5:strain}, \eq{eq:wave:ph}, and \eq{eq:CSW:el}, their amplitudes are fixed by
\beqn
{\hat M} V_k = 0, 
\qquad 
V_k = \Vector{u^z_k}{\rho_{5,k}}, 
\quad
\label{eq:MV}
\eeqn
where the mixing matrix is
\beqn
{\hat M} = \Matrix{\omega^2 - v_s^2 k^2}{- i \kappa b v_\CSW v_s^2 k}{\frac{i \theta b^2}{6 \pi^2} k \omega }{\omega - v_\CSW k + \frac{i}{\tau_5}}.
\label{eq:M}
\eeqn

The various branches of the energy dispersion $\omega = \omega(k)$ are determined by the requirement that the determinant of the matrix Eq.~\eq{eq:M} vanishes:
\beqn
\bigl[(\omega+ i \nu_s)^2 {-} v_s^2 k^2 \bigr] \left( \omega + i \nu_5 {-} v_\CSW k \right) - v_p^2 \omega k^2  = 0, \qquad
\label{eq:mixing:nu}
\eeqn
where we have added a sound attenuation rate ~$\nu_s = 1 / \tau_s$ which accounts any source of the (ultra)sound dissipation  other than the scattering of the chiral and ordinary phonons (for simplicity we neglect possible anisotropies of the sound attenuation).
According to Eq. \eqref{eq:mixing:nu}, the strength of the mixing between the ordinary and chiral phonons is given by the parameter
\beqn
v_p = v_s \sqrt{\frac{\kappa \theta v_\CSW b^3}{6 \pi^2}}, 
\label{eq:v:p}
\eeqn
which has the dimension of velocity (m/s).

The inter-valley scattering rate $\nu_5 = 1 / \tau_5$, the velocity of the longitudinal sound $v_s$, and the sound attenuation rate~$\nu_s$ are, basically, strain-independent parameters which are determined by the crystal and electronic  structure of a particular Weyl semimetal. On the contrary, the velocity $v_\CSW$ of the chiral phonons and the coupling $v_p^2$ between the chiral phonons and the ordinary phonons are strain-dependent quantities linearly proportional to the twisting angle~$\theta$. In the long-wavelength limit, $k \to 0$, the solutions of Eq.~\eq{eq:mixing:nu} decouple into three branches. The first branch corresponds to the chiral sound which carries fluctuations of the axial charge coupled to the elastic deformations (elastic sound waves):
\beqn
\omega = v_\CSW k - i \nu_5,
\label{eq:CSW:IR}
\eeqn
and the amplitude~\eq{eq:MV}
\beqn
V^\CSW_k =\frac{1}{N_\CSW}  \Vector{ - i b k \kappa v_\CSW v_s^2\vspace{1mm}}{  (\nu_5 - \nu_s)^2} \bigl( 1+ O(k)\bigr),
\label{eq:Vk:CSW}
\eeqn
where $N_\CSW$ is a normalization factor. It is important to note that the propagating ($k \neq 0$) chiral sound is always accompanied by the usual longitudinal sound wave due to the nonzero upper component in the amplitude~\eq{eq:Vk:CSW}. The chiral sound  will in turn affect the dispersion relation of ordinary phonons, a direct measurable effect.

The long-wavelength solutions of Eq.~\eq{eq:mixing:nu} include also two branches of the standard acoustic phonons which propagate in  both senses along ${\bs B}_5$ with the dispersion:
\beqn
\omega = \pm v_{\mathrm{ac}} k - i \nu_s.
\label{eq:mixed:IR}
\eeqn
The velocity of the acoustic phonons,
\beqn
v_{\mathrm{ac}} = \sqrt{v_s^2 - \frac{\nu_s}{\nu_5 - \nu_s} v_p^2 },
\label{eq:v:mix}
\eeqn
is affected by the presence of the chiral phonons via the ``mixing velocity'' $v_p = v_p(\theta)$, given in Eq.~\eq{eq:v:p}. 

The amplitude~\eq{eq:MV} of the acoustic branches~\eq{eq:mixed:IR} is
\beqn
V^{\mathrm{ac}}_k =\frac{1}{N_{\mathrm{ac}}}  \Vector{6 \pi^2 (\nu_5 - \nu_s)\vspace{1mm}}{ i \theta b^2 \nu_s k} \bigl( 1+ O(k)\bigr),
\label{eq:Vk}
\eeqn
where $N$ is an appropriate normalization. The mixing of the acoustic phonons and chiral density fluctuations vanishes in the strict long-wavelength limit $k \to 0$. In this limit, the dissipation  of the chiral sound wave~\eq{eq:CSW:IR} and the two acoustic branches~\eq{eq:mixed:IR}  is 
given by the inter--valley $\nu_5$ and sound $\nu_s$ dissipation rates, respectively. 
\begin{figure}[htb]
\begin{center}
\includegraphics[scale=0.55,clip=true]{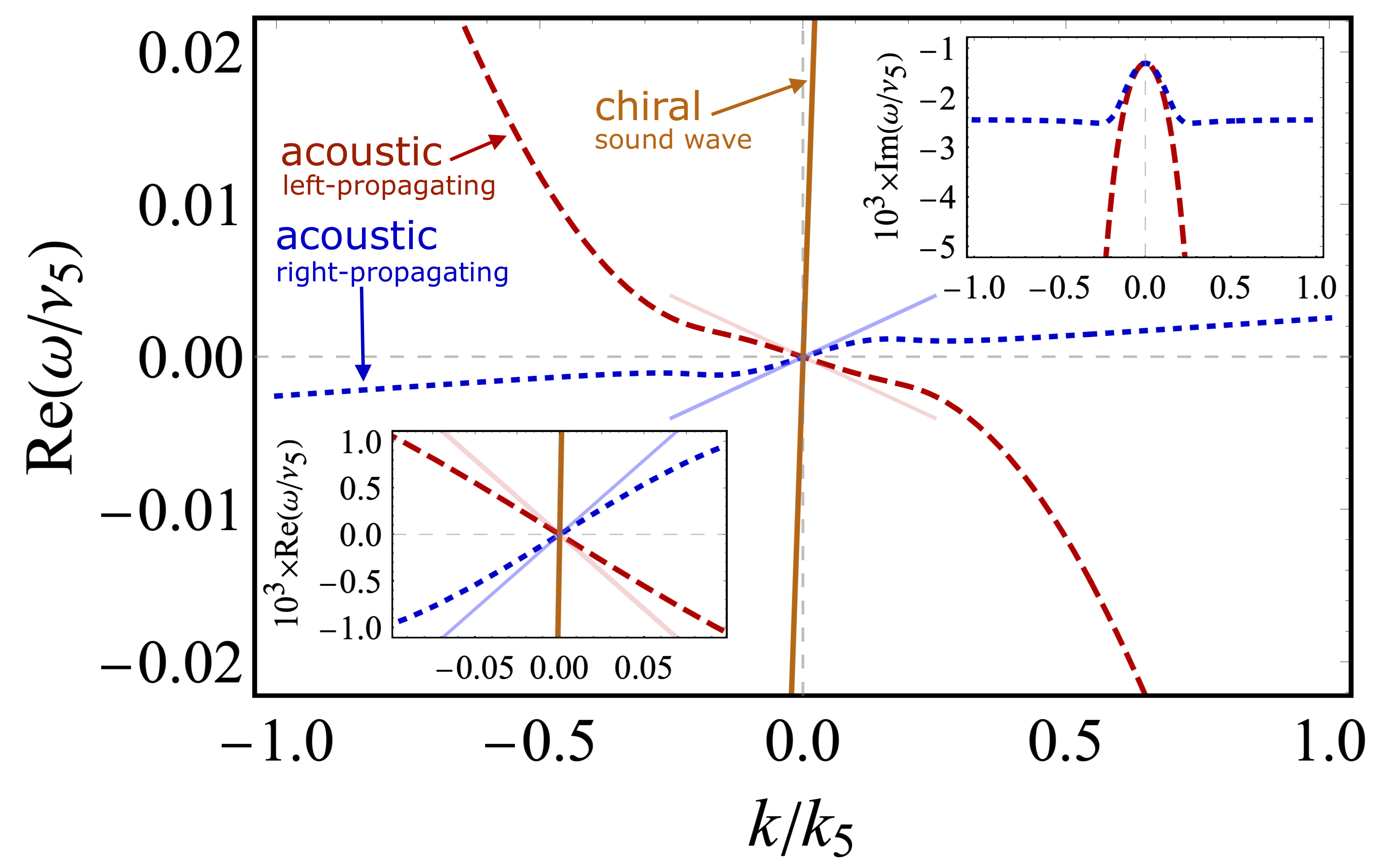}
\end{center}
\vskip -5mm
\caption{The dispersions $\omega = \omega(k)$ of the chiral phonon (the so\-lid orange line) and two branches of the acoustic phonons (the red dashed and blue dotted lines) for the Weyl semimetal TaAs subjected to a strong strain ($v_\CSW = v_F$ in the quantum limit) with a weak acoustic-chiral phonon mixing $v_p^2 = 0.1 v_F^2$. Other parameters are given in Table~\ref{tbl:parameters}. The momentum $k$ is plotted in units of $k_5 = \nu_5/v_F$. A zooming of the long-wavelength region of the real part of the dispersions in shown in the left inset. 
The right inset shows the imaginary part of these dispersions.  The light solid lines depict the real parts of the dispersions for an unstrained ($\theta=0$) semimetal. 
} 
\label{fig:dispersions}
\end{figure}

A numerical analysis of the dispersions of the three phonon branches made with the standard parameters of TaAs is shown in Fig. \ref{fig:dispersions}. 
There are two main regimes of propagation set by the scale of the momentum~$k$.  

At low momenta $k$, the linear phonon branches~\eq{eq:v:mix} exist at sufficiently low strains $\theta$. These branches disappear at the critical strain defined by the equation $v_\CSW \theta_c = 6 \pi^2 (\nu_5 - \nu_s)/(\kappa b^3 \nu_s)$ above which the long-wavelength acoustic phonons acquire a quadratic dispersion, $\omega \sim k^2$. The dispersion of the chiral phonons remains linear. 

At larger momentum $k$, the dispersions of all three branches, the chiral-like ($\ell=0$) and the acoustic-like  ($\ell=\pm$), 
become linear again:
\beqn
\omega = v_\ell k - i \nu_\ell,  \qquad \ell = 0,\pm,
\eeqn
where the velocities $v_\ell$ are the roots of the equation:
\beqn
v^3 - v_\CSW v^2 - \left( v_p^2 + v_s^2 \right) v + v_\CSW v_s^2 = 0. 
\eeqn 
The velocities $v_\ell$ are, in general, different if $v_p \neq 0$.

If the mixing between the chiral and acoustic phonons is small ($v_p \ll v_s$ and $v_p \ll v_\CSW \ll v_s$ with $v_s \neq v_\CSW$), the velocities of all three branches are as follows:
\beqs
\beqn
v_0 & = & v_\CSW + \frac{v_\CSW}{v_\CSW^2 - v_s^2} v_p^2 + O(v_p^4)\,, \\
v_\pm & = & \pm v_s - \frac{1}{2(v_\CSW \mp v_s)} v_p^2 + O(v_p^4)\,.
\label{eq:vpm}
\eeqn
\label{eq:vs:small}
\eeqs 
Equations~\eq{eq:vs:small} imply that the velocities of the acoustic phonons in opposite directions of the rod after mixing with the CSW will differ by an amount proportional to the magnitude and the sign of the twist~$\theta$:
\beqn
\delta v = |v_+| - |v_-| \simeq v_p^2/v_s.
\label{eq:delta:v}
\eeqn
Here we used Eqs.~\eq{eq:wave:ph} and \eq{eq:v:p}, and assumed that $v_\CSW \ll v_s$, in accordance with an experimentally relevant case discussed below. As we can see in Fig.~\ref{fig:dispersions}, in the long-wavelength regime the velocity split becomes less pronounced, while the counter-propagating acoustic waves appear to have unequal dissipation rates due to the different coupling to the chiral phonons. 

The magnitude of the typical parameters for the Weyl semimetal TaAs is given in Table~\ref{tbl:parameters}
\cite{Lv15,LXH15,ZXetal16}. 
\begin{table}[htb]
\begin{tabular}{|c|c|c|c|c|c|}
\hline 
\multicolumn{6}{|c|}{TaAs} \\
\hline 
\multicolumn{2}{|c|}{quasiparticles} & \multicolumn{4}{c|}{phonons} \\
\hline 
$v_F$ (m/s) & $\nu_5$ (1/s) & $v_s$ (m/s) & $\nu_s$ (1/s) & $v_s/v_F$ & $\nu_s / \nu_5$ \\
\hline 
$3 {\times} 10^5$ & $2 {\times} 10^9$ & $4.8 {\times} 10^3$ & $2.6 {\times} 10^6$ &  $1.6 {\times} 10^{-2}$ & $1.3 {\times} 10^{-3}$ \\
\hline 
\end{tabular}
\caption{Typical reference parameters for the TaAs semimetal: shown are the Fermi velocity $v_F \equiv v_z$ in the W1 pocket along the $z$ axis and the $v_s \equiv v_{zz}$ velocity for the longitudinally polarized ultrasound along the same $z$ axis.}
\label{tbl:parameters}
\end{table}

Taking the separation $|2 {\bs b}| \simeq 0.3$\,\AA${}^{-1}$ for the Weyl nodes of TaAs, we estimate $B_5 \simeq 17\,\mbox{mT}$  for a $1^{\circ}$ twist of $L=1\,\mu\mbox{m}$ long wire corresponding to the torsion angle $\theta = \frac{2 \pi}{360} \frac{1}{L} \simeq 1.7 \times 10^4\,\mbox{m}^{-1}$. 
In a low temperature regime, at $\mu \simeq 26$\,meV~\cite{ref:TaAs:mu},  the CSW propagates with the velocity~\eq{eq:CSW:el}  $v_\CSW \simeq  225$\,m/s. Taking into account that the Gruneisen parameter $\beta$ is of order one in most materials~\cite{IL09}, and given the values of the shear ($G=54$\,GPa) and bulk ($B=189$\,GPa) moduli for TaAs~\cite{ref:structural}, one finds that the value $v_p \simeq 10^{-3}$\,m/s for the acoustic-chiral mixing parameter~\eq{eq:v:p}. Consequently, we get a value for the splitting of the left/right-propagating modes $\delta v \sim 3 \times 10^{-9}$ m/s at a low-temperature regime. While this splitting itself is too small for experimental detection, the chiral sound wave can be detected experimentally through its slow-velocity imprint ($v_\CSW \simeq  225$\,m/s) in the sound spectrum due to the coupling of the chiral-wave amplitude to the elastic sound sector~\eq{eq:Vk:CSW}.

We did not consider finite geometry effects which can alter slightly the results. A very complete discussion of this issue as well as a proposal of sound attenuation related to $E_5$ and $B_5$ effects can be found in Ref.~\cite{PCF16}.  
\paragraph{Summary and perspectives.}
Based on the same physics that gives rise to the chiral magnetic wave \cite{KY11}, but driven by axial elastic gauge fields, we have found a  unidirectional chiral sound wave (CSW) which  propagates longer and dissipates slower than the chiral magnetic wave  making better prospects for its experimental detection. The key difference is that our CSW does not hybridize with plasmons to linear order in the derivative expansion and avoids the over-damping predicted in \cite{SRG18}. Instead, it does hybridize with standard acoustic phonons, what provides  the way to its experimental detection. The CSW propagates in the direction of the elastic magnetic field in one direction only. Such an experiment will provide a confirmation of the  AAA contribution to the chiral anomaly, and an additional evidence for the presence of elastic axial gauge fields in Weyl semimetals. 

Other interesting proposals of chiral waves in Weyl semimetals (normally in magnetic or pseudomagnetic  fields) can be found in Refs.~\cite{PCF16,GMetal17b,Dai19}. Acoustic phonons running with different velocities along the opposite directions of a magnetic field in a chiral (not WSM) material have been described in \cite{NZetal19}. 

The search for evidences of the chiral anomaly in WSMs away from the standard magneto-electric measurements has become  a very active field in the area \cite{SZetal16,RLG17,HQetal18,NZetal19,HZK19}. Previous proposals of using standard phonons are dimmed by their hybridization with collective electronic excitations while some materials, as TaAs, do not support pseudoscalar phonons \cite{SZetal16,RLG17}. Our  mechanism will occur generically in all materials with well separated Weyl nodes. The chiral wave analyzed in this work will perhaps be one of the cleanest evidences for the chiral anomaly coming from the less common AAA diagram.

More exotic Weyl metamaterials made with optical, acoustic, or magnon lattices \cite{HNetal16,ZMetal18,FBB15,HGetal19} appeared recently in the material sciences. They may form new platforms for chirality--based quantum computing~\cite{Dima19}. An interesting perspective is to see if the proposed chiral sound wave emerge in the Weyl materials where elastic gauge fields are also present~\cite{FV18,PSetal19}.

\begin{acknowledgments}
The authors are grateful to Alberto Cortijo, Yago Ferreiros, Karl Landsteiner, Imam Makhfudz, and Igor Shovkovy for discussions. This work was conceived during the Work\-shop on Weyl Metals at the Instituto de F\'isica Te\'orica de Madrid, February 2019. This paper was partially supported by Spanish MECD grant FIS2014-57432-P, the Comunidad de Madrid MAD2D-CM Program (S2013/MIT-3007),  Grant 3.6261.2017/8.9 of the Ministry of Science and Higher Education of Russia, and Spanish--French mobility project PIC2016FR6/PICS07480.
\end{acknowledgments}

\end{document}